\documentclass[conference,twoside,print,9pt]{ieeecolor}
\usepackage{generic}
\usepackage{cite}
\usepackage{amsmath,amssymb,amsfonts,mathrsfs}
\usepackage{algorithmic}
\usepackage{graphicx}
\usepackage{algorithm,algorithmic}
\usepackage{hyperref}
\usepackage{ulem}   
\hypersetup{hidelinks=true}
\usepackage{textcomp}
\usepackage{relsize}
\usepackage[most]{tcolorbox}
\def\BibTeX{{\rm B\kern-.05em{\sc i\kern-.025em b}\kern-.08em
    T\kern-.1667em\lower.7ex\hbox{E}\kern-.125emX}}
\usepackage{tikz,pgfplots}
\pgfplotsset{compat=1.16}
\usetikzlibrary{arrows.meta, shapes.geometric, shadows, positioning, calc, backgrounds}

\usepgfplotslibrary{patchplots}
\usepgfplotslibrary{fillbetween}
\pgfplotsset{%
     layers/standard/.define layer set={%
         background,axis background,axis grid,axis ticks,axis lines,axis tick labels,pre main,main,axis descriptions,axis foreground%
     }{
         grid style={/pgfplots/on layer=axis grid},%
         tick style={/pgfplots/on layer=axis ticks},%
         axis line style={/pgfplots/on layer=axis lines},%
         label style={/pgfplots/on layer=axis descriptions},%
         title style={/pgfplots/on layer=axis descriptions},%
         colorbar style={/pgfplots/on layer=axis descriptions},%
         ticklabel style={/pgfplots/on layer=axis tick labels},%
         axis background@ style={/pgfplots/on layer=axis background},%
         3d box foreground style={/pgfplots/on layer=axis foreground},%
     },
 }

\newtheorem{proposition}{Proposition}
\newtheorem{assumption}{Assumption}
\newtcbtheorem{corollary}{Corollary}{
  enhanced,
  attach title to upper={\ ---\ }, 
  rounded corners,
  colback=white,
  coltitle=black,
  colbacktitle=white!90!black,
  colframe=gray!75!black,
  fonttitle=\bfseries
}{corr}

\newtcbtheorem{theorem}{Theorem}{
  enhanced,
  attach title to upper={\ ---\ }, 
  rounded corners,
  colback=white,
  coltitle=black,
  colbacktitle=white!90!black,
  colframe=gray!75!black,
  fonttitle=\bfseries
}{thm}
\newtheorem{remark}{Remark}
\newtheorem{example}{Example}%
\newtheorem{definition}{Definition}%
\def\IEEEmembership#1{\textit{#1}}
\definecolor{nblue}{cmyk}{0,0,0,1}

\newcommand{\thank}[1]{
    \begingroup
    \renewcommand\thefootnote{}\footnote{#1}%
    \addtocounter{footnote}{-1}%
    \endgroup    
}%

\begin{document}
\title{Counterclockwise Dissipativity, Potential Games and Evolutionary Nash Equilibrium Learning}
\author{Nuno C. Martins, \IEEEmembership{Senior Member, IEEE}, Jair Cert\'{o}rio, \IEEEmembership{Student Member, IEEE}, \\ and Matthew S. Hankins, \IEEEmembership{Student Member, IEEE}
}

\maketitle

\begin{abstract}
We use system-theoretic passivity methods to study evolutionary Nash equilibria learning in large populations of agents engaged in strategic, non-cooperative interactions. The agents follow learning rules (rules for short) that capture their strategic preferences and a payoff mechanism ascribes payoffs to the available strategies. The population's aggregate strategic profile is the state of an associated evolutionary dynamical system. Evolutionary Nash equilibrium learning refers to the convergence of this state to the Nash equilibria set of the payoff mechanism. Most approaches consider memoryless payoff mechanisms, such as potential games. Recently, methods using $\delta$-passivity and equilibrium independent passivity (EIP) have introduced dynamic payoff mechanisms. However, $\delta$-passivity does not hold when agents follow rules exhibiting ``imitation" behavior, such as in replicator dynamics. Conversely, EIP applies to the replicator dynamics but not to $\delta$-passive rules. We address this gap using counterclockwise dissipativity (CCW). First, we prove that continuous memoryless payoff mechanisms are CCW if and only if they are potential games. Subsequently, under (possibly dynamic) CCW payoff mechanisms, we establish evolutionary Nash equilibrium learning for any rule within a convex cone spanned by imitation rules and continuous $\delta$-passive rules. %
\end{abstract}


\thank{This work was supported by 
AFOSR Grant FA9550-23-1-0467 and 
NSF Grants 2139713 and  
2135561.}
\thank{Nuno C. Martins, Jair Cert\'{o}rio and Matthew S. Hankins are with the Department of Electrical and Computer Engineering and the Institute For Systems Research at the University of Maryland, College Park \\ (e-mail: nmartins@umd.edu, certorio@umd.edu, msh@umd.edu). }

\section{Introduction}
\label{sec:introduction}
\IEEEPARstart{R}{ecent} developments in population games and evolutionary dynamics~\cite{Sandholm2010Population-Game,Sandholm2015Handbook-of-gam} have introduced new systematic methods to model and analyze the dynamics of strategic, non-cooperative interactions among large populations of learning agents. Learning rules (\uline{rules for short}), also referred to as strategy revision protocols, describe how agents update their strategies based on the payoffs these strategies yield. The rules, which typically seek strategies with higher payoffs, can be programmed into artificial agents or represent the preferences or bounded rationality of humans and other natural agents.

The agents are nondescript and the population's aggregate strategic profile is represented as a \uline{population state} vector whose entries are the proportions of the population selecting the available strategies. As agents revise their strategies, the population state changes over time according to an evolutionary dynamic model resulting from the rule in effect.  A payoff mechanism generates the strategies' payoffs and might abstractly represent the specifics of the agents' interaction environments, such as in congestion games~\cite{Beckmann1956Studies-in-the-}, or be designed by a coordinator to nudge the population state towards desired equilibria~\cite{Quijano2017The-role-of-pop,Martins2023Epidemic-popula}.
This framework has been suitable for distributed optimization~\cite{Barreiro-Gomez2016Constrained-dis} and engineering systems~\cite{Tembine2010Evolutionary-ga,Obando2014Building-temper}. \\

We seek to understand how populations that follow certain rules achieve and maintain Nash equilibria, a process we call \emph{evolutionary Nash equilibrium learning} to distinguish it from \emph{Nash equilibrium seeking}~\cite{Frihauf_2012aa} that is used more generally when the learning process is not necessarily expressible using evolutionary dynamics, such as in~\cite{Gadjov_2019aa}. Specifically, we say that a population evolutionarily learns Nash equilibria when its population state converges to the Nash equilibria set appropriately defined for the payoff mechanism. Establishing this property is crucial because, when it holds, the Nash equilibria set can be used to predict the long-term evolution of the population state. In certain applications where one has the authority to design the payoff mechanism, the Nash equilibria have inherent optimality properties for potential games~\cite[\S3~and~\S5]{Sandholm2001Potential-games} and may even be selected to satisfy performance requirements~\cite{Martins2023Epidemic-popula,Certorio2022Epidemic-Popula}.

\subsection{Background On Passivity Approaches}
\label{subsec:PassivityApproaches}

As evident from the synopsis~\cite{Sandholm2015Handbook-of-gam}, most research on evolutionary learning of Nash equilibria has focused on memoryless payoff mechanisms until recently. Of particular relevance is the focus on memoryless payoff mechanisms that are potential games~\cite{Sandholm2001Potential-games}. Following this, contractive games were introduced~\cite{Hofbauer2009Stable-games-an}, which have the desirable properties of concave potential games without necessarily being potential.

A novel approach utilizing system theoretic passivity methods extended the results in~\cite{Hofbauer2009Stable-games-an} to include dynamic payoff mechanisms~\cite{Fox2013Population-Game} that are the dynamic analogs of contractive games. This extension was achieved through the introduction of the concept of $\delta$-passivity, which was further explored in~\cite{Park2019From-Population, Arcak2020Dissipativity-T} and applied in various design applications~\cite{Martins2023Epidemic-popula, Certorio2022Epidemic-Popula}. Recent work established $\delta$-passivity for convex cones of hybrid rules, formed as conic combinations of rules from canonical classes known to satisfy $\delta$-passivity~\cite{Certorio_2024aa}. \\

\textbf{An important gap:} Despite the significant success with $\delta$-passivity approaches, \cite[Proposition~III.5]{Park2018Passivity-and-e} proved that the replicator rule associated with the well-known replicator dynamics is not $\delta$-passive. More generally, no rule exhibiting ``imitation'' behavior has been shown to be $\delta$-passive. Conversely, \cite{Mabrok2021Passivity-Analy} proved that the replicator rule is equilibrium independent passive (EIP), a passivity notion not known to hold for the $\delta$-passive rules in~\cite{Certorio_2024aa}.

\subsection{Main Objective}
\label{subsec:MainObjective}

This paper aims to address the gap identified above by developing a system-theoretic passivity approach that is compatible with a broader set of rules, including both the continuous hybrid $\delta$-passive rules described in \cite{Certorio_2024aa} and rules exhibiting imitation behavior, such as the replicator rule. Specifically, we seek an approach that can accommodate conic combinations of imitation-based rules  and the continuous $\delta$-passive rules considered in \cite{Certorio_2024aa}. This goal is particularly important when we are given or intend to design a payoff mechanism and seek evolutionary Nash equilibrium learning guarantees without precise knowledge of the population's rule. The broader the class of rules that our passivity method can handle, the more general and robust the guarantees we can provide.

\subsection{Contributions And Limitations Of Our Approach}
\label{subsec:ContribAndLim}

Motivated by the observation that most rules, including the replicator rule and those in \cite{Certorio_2024aa}, satisfy a property known as \emph{positive correlation}, we bring to bear a closely related system-theoretic concept referred to as \emph{counterclockwise (CCW) dissipativity} proposed in \cite{Angeli2006Systems-With-Co} to study the stability and robustness of feedback nonlinear systems. A similar notion based on negative imaginary inequalities has been originally proposed in~\cite{Lanzon2008Stability-Robus} for linear time invariant systems with multiple inputs and outputs. \\

The following are our contributions.
\begin{itemize}
    \item[{\bf C1}] Theorem~\ref{thm:CCWGameIsPotential} in \S\ref{sec:MainResult}, states that a continuous \uline{memoryless payoff mechanism is a potential game if and only if it is CCW}. Hence, the set of CCW payoff mechanisms can be rightly viewed as an extension of potential games in the same way that the set of $\delta$-passive payoff mechanisms includes memoryless payoff mechanisms that are contractive games as a particular case.
    \item[{\bf C2}] For an expanded set of hybrid rules we will rigorously define in~\S\ref{subsec:HybridRules}, Theorem~\ref{thm:Main} in \S\ref{sec:MainResult} states that the population state converges to an equilibrium set when the payoff mechanism is CCW. Furthermore, this equilibrium set corresponds to the Nash equilibria set of the so-called stationary game appropriately defined for the given payoff mechanism, thereby establishing the evolutionary learning of the Nash equilibrium set.
    \item[{\bf C3}] The aforementioned expanded hybrid rule set for which Theorem~\ref{thm:Main} holds is a convex cone that bridges the gap identified in~\S\ref{subsec:PassivityApproaches}. 
\end{itemize}
\vspace{.1 in}
Despite these results, our work does not supersede existing system-theoretic passivity approaches in all cases. The following are limitations of our work and mitigating factors.
\begin{itemize}
    \item[{\bf L1}] A payoff mechanism that is not CCW may still possess other passivity properties that would make it stabilizing for a $\delta$-passive or replicator rule. We provide a detailed comparison in \S\ref{subsec:ComparisonPassiveApproaches}.
    \item[{\bf L2}] In the cone of hybrid rules we consider, the replicator rule, as well as rules featuring imitation behavior, cannot appear in isolation as in~\cite{Mabrok2021Passivity-Analy} but are always combined with at least one ``non-imitation" rule. The weight of the non-imitation rules can be arbitrarily small, so long as it is positive, which allows us to obtain approximations of ``pure" imitation rules to an arbitrary degree of accuracy. 
    \item[{\bf L3}] For simplicity, our framework assumes that the hybrid rules cannot be set-valued and must be continuous to avoid having to consider nonstandard notions of positive correlation, differential inclusions, and associated Filippov or Carathéodory notions of solution \cite{Cortes_2008aa}. For this reason, unlike \cite{Certorio_2024aa}, the rules considered here cannot involve the so-called best response rule. Fortunately, as evidenced by Example~\ref{ex:ApprxBestResponse} presented later in \S\ref{subsec:Canonicalrules}, the cone of hybrid rules we consider includes approximations of the best response rule to an arbitrary degree of accuracy.
\end{itemize}

\subsection{Outline of the Paper}

After the Introduction, 
\S\ref{sec:FrameworkandFormulation} describes the evolutionary dynamics model and learning rules used in our framework.
We introduce the concepts of positive correlation, Nash stationarity, and hybrid learning rules in  \S\ref{sec:NSandPC}.
In \S\ref{sec:MainResult} we describe CCW payoff mechanisms, their connection to potential games, and our main result, that populations that use positively correlated rules will converge to a Nash equilibrium of a CCW payoff mechanisms.   
We illustrate our convergence result by simulating a CCW payoff dynamic model in \S\ref{sec:Simulation}, and we summarize our findings in \S\ref{sec:Conclusions}.

\paragraph*{\bf Notation} We use $A'$ to denote the transpose conjugate of a matrix $A$ in $\mathbb{C}^{n \times m}$. We use $A \succeq 0$ and $A \succ 0$  to indicate, respectively, that a square matrix $A$ is positive semi-definite and positive definite. Additional notation and definitions will be introduced as needed.

\section{Framework and Problem Formulation}
\label{sec:FrameworkandFormulation}

Our framework models the noncooperative strategic interactions of a large number of agents. For simplicity, we assume that all the agents belong to one population characterized by a finite set of available strategies $\{1,\ldots,n\}$.
Each agent follows one strategy at a time, which the agent can revise repeatedly. A payoff vector $p(t)$ in $\mathbb{R}^n$, whose $i$-th entry $p_i(t)$ quantifies the net rewards of the strategy $i$ at time $t$, influences the revision process, as typically the agents seek strategies with higher payoffs. The agents are nondescript and the so-called population state vector $x(t)$ in 
${\mathbb{X}:= \Big \{x \in [0,1]^n \ \big | \ \textstyle\sum_{\ell=1}^n x_\ell =1 \Big \}}$
approximates the population's aggregate strategic choices in the large population limit~\cite{Sandholm2003Evolution-and-e}. Namely, $x_i(t)$ approximates the proportion of the population's agents following strategy $i$ at time $t$. 

\subsection{Learning Rules And Evolutionary Dynamics}
The revision process causes $x$ to vary over time. Following the standard approach in~\cite[\S4.1.2]{Sandholm2010Population-Game}, the following {\it evolutionary dynamics model} {\bf (EDM)} governs the dynamics of~$x$:
\begin{subequations}
\begin{equation}\tag{EDMa} \label{eq:EDM-DEF} \dot {x}(t) = \mathcal{V} ( x(t),p(t) ), \quad t\geq 0, 
\end{equation}
where the $i$-th component of $\mathcal{V}$ accounts for the net flow of agents switching to strategy $i$ according to 
\begin{equation} \tag{EDMb}
\underbrace{\mathcal{V}_i ( x(t),p(t) )}_{\text{\small net flow into $i$}} := \sum_{j=1}^n \underbrace{x_j \mathcal{T}_{ji}(x,p)}_{\text{\small flow} \ j\rightarrow i} - \underbrace{x_i \mathcal{T}_{ij}(x,p)}_{\text{\small flow} \ i\rightarrow j}.
\end{equation}
\end{subequations} 

A Lipschitz continuous map $\mathcal{T}: \mathbb{X} \times \mathbb{R}^{n} \to \mathbb{R}_{\geq 0}^{n \times n}$,  %
referred to as {\it learning rule}\footnote{Learning rules are also denoted in the literature as strategy revision protocols.} (\uline{rule for short}), models the agents' strategy revision preferences. In \cite[Part~II]{Sandholm2010Population-Game} and \cite[\S 13.3-13.5]{Sandholm2015Handbook-of-gam} there is a comprehensive discussion on rule types and the classes of bounded rationality behaviors they model. 

\subsection{Payoff Mechanism And Solutions}
It suffices to consider Lipschitz continuous (w.r.t. time) population state trajectories $\mathbf{x}$, leading to the set
\begin{equation*}
\mathscr{X}:=\Big \{ \mathbf{x}:[0,\infty]\to \mathbb{X} \ \big | \  \text{$\mathbf{x}$ { is Lipschitz continuous} } \Big \}.
\end{equation*} It also suffices to consider Lipschitz continuous (w.r.t time) payoff trajectories $\mathbf{p}$, leading to the set 
\begin{equation*}
\mathscr{P}:=\Big \{ \mathbf{p}:[0,\infty]\to \mathbb{R}^n \ \big | \  \text{$\mathbf{p}$ { is Lipschitz continuous}} \Big \}.
\end{equation*}

We can now define the payoff mechanisms we will use. \\
\begin{definition} A \underline{payoff mechanism} $\mathfrak{F}:\mathscr{X} \to \mathscr{P}$ is a map generating an output $\mathbf{p}$ in $\mathscr{P}$ for each input $\mathbf{x}$ in $\mathscr{X}$. 
\end{definition}
\vspace{.1 in} 
 When discussing a payoff mechanism, we denote its input as $\mathbf{x}$ and its output as $\mathbf{p}$. The roles for $\mathbf{x}$ and $\mathbf{p}$ are reversed for an (EDM), where the former is the output and the latter the input. In cases when the payoff mechanism is a dynamical system with an internal state $\mathbf{q}$, without loss of generality, to simplify notation we consider a zero initial condition $q(0)=0$. \\

\begin{figure}
\begin{center}
\begin{tikzpicture}[
                scale = 1,
                transform shape,
                node distance=0.5cm,
                block/.style={rectangle, draw, rounded corners, minimum width=2.3cm, minimum height=1.2cm, align=center},
                arrow/.style={->, >=stealth, ultra thick}
                ]

                \node[block, fill=teal!8] (strat) at (0,0) {\large $\dot{x} = \mathcal{V}(x,p)$\\{\large (EDM)}};
                \node[block, fill=orange!8, below=of strat] (payoff) {\Large $\mathfrak{F}:\mathbf{x} \mapsto \mathbf{p}$};

                \draw[-{Latex[length=3mm, width=2mm]}, thick] (payoff.west) -- ++(-1,0) |- (strat.west) node[midway, left] {\Large $\mathbf{p}$};
                \draw[-{Latex[length=3mm, width=2mm]}, thick] (strat.east) -- ++(1,0) |- (payoff.east) node[midway, right] {\Large $\mathbf{x}$};
                
\end{tikzpicture}

\end{center}
\caption{Interconnection of (EDM) and payoff mechanism $\mathfrak{F}$.}
\label{fig:closedloop}
\end{figure}
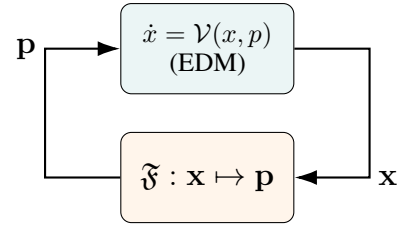

\begin{assumption}
\label{ass:PayoffMech} Any payoff mechanism $\mathfrak{F}$ in our analysis is considered to satisfy the following assumptions:
\begin{enumerate}%
    \item For any initial condition $x(0)$ in $\mathbb{X}$, the closed-loop system in Fig.~\ref{fig:closedloop} formed by interconnecting $\mathfrak{F}$ in feedback with any (EDM) has a unique solution pair $(\mathbf{x},\mathbf{p})$. In addition, $\mathbf{x}$ and $\mathbf{p}$ are, respectively, in $\mathscr{X}$ and $\mathscr{P}$.
    \item There is a Lipschitz continuous map $\mathcal{F}_\mathfrak{F}:\mathbb{X} \to \mathbb{R}^n$ such that for every $\mathbf{x}$ in $\mathscr{X}$, the following holds:
    $$ \lim_{t \rightarrow \infty} \| \dot{x}(t) \|=0 \implies \lim_{t \rightarrow \infty} \big \| p(t)-\mathcal{F}_{\mathfrak{F}} \big(x(t)\big) \big \| = 0.$$ We refer to $\mathcal{F}_{\mathfrak{F}}$ as the \uline{stationary game} of $\mathfrak{F}$.
    \item There is a constant $\beta_{\mathfrak{F}}$ such that the following holds:
    $$ \sup_{t \geq 0} \| p(t) \|_\infty \leq \beta_{\mathfrak{F}} < \infty, \ \mathfrak{F}:\mathbf{x} \mapsto \mathbf{p}, \ \mathbf{x} \in \mathscr{X}.$$
\end{enumerate}
\end{assumption}
\vspace{.1 in}
\begin{remark} We observe that Rademacher's Theorem implies that all $\mathbf{x}$ in $\mathscr{X}$ and $\mathbf{p}$ in $\mathscr{P}$ are differentiable for almost all time $t \geq 0$. We will be implicitly using this fact when writing integrals with respect to time of integrands involving functions of $\dot{\mathbf{x}}$ and $\dot{\mathbf{p}}$. For this reason, henceforth, all integrals are in the sense of Lebesgue. Furthermore, if $(\mathbf{x},\mathbf{p})$ is a solution pair as specified in Assumption~\ref{ass:PayoffMech}.1, then $\dot{x}(t)$ exists for all $t \geq 0$ because (EDM) must hold for all $t \geq 0$.
\end{remark}
\vspace{.1 in}
\subsubsection{Games and Potential Games}
We proceed to define two classes of payoff mechanisms satisfying Assumption~\ref{ass:PayoffMech} that we will use in our analysis and to illustrate our results later on. \\

\begin{definition}{\bf [Game and Potential Game]} \label{def:game} A payoff mechanism $\mathfrak{F}$ is referred to as {\it a game} when it is a memoryless map ${\mathcal{F}:x(t) \mapsto p(t)}$, where $\mathcal{F}:\mathbb{X} \to \mathbb{R}^n$ is Lipschitz continuous. In this case, $\mathcal{F}$ and $\mathcal{F}_\mathfrak{F}$ are identical. A game $\mathcal{F}$ is a \uline{potential game} when there is a potential $f:\mathbb{X} \to \mathbb{R}_{\geq 0}$ satisfying\footnote{See~\cite{Sandholm2001Potential-games} and references therein for a detailed discussion.}
\begin{equation}
\label{eq:PotentialGame}
f\big ( x(T) \big) - f\big ( x(0) \big) = \int_0^T \dot{x}'(t)\mathcal{F}\big ( x(t) \big) dt, \ T>0, \ \mathbf{x} \in \mathscr{X}.
\end{equation}
\end{definition} 
\vspace{.1 in}

If $\hat{f}:U \to \mathbb{R}$ is a continuously differentiable extension of $f$ to an open set $U$ containing $\mathbb{X}$ for a potential game $\mathcal{F}$ then $(v-x)'\nabla \hat{f}(x) = (v-x)'\mathcal{F}(x)$, for all $x$ and $v$ in $\mathbb{X}$. If $\mathcal{F}$ admits a continuously differentiable extension $\hat{\mathcal{F}}:U \to \mathbb{R}^n$ to an open set $U$ containing $\mathbb{X}$ then $\mathcal{F}$ is a potential game if and only if $(w-x)'\big( J \hat{\mathcal{F}} (x)-J \hat{\mathcal{F}}'(x))(v-x)=0$ for all $w$, $v$ and $x$ in $\mathbb{X}$, where $J\hat{\mathcal{F}}(x)$ is the Jacobian of $\hat{\mathcal{F}}$ evaluated at $x$. If, in addition, $(v-x)'\big( J \hat{\mathcal{F}} (x))(v-x) \leq 0$ for all $x$ and $v$ in $\mathbb{X}$ then $f$ is concave and $\mathcal{F}$ is said to be contractive\cite[13.7.2]{Sandholm2015Handbook-of-gam}. Hence, $\mathcal{F}(x)=Ax+b$ is a potential game for any $b$ in $\mathbb{R}^n$ and $A=A'$ in $\mathbb{R}^{n \times n}$, with $\mathcal{F}$ contractive when $A \preceq 0$. The congestion game\footnote{See also~\cite{Beckmann1956Studies-in-the-}.} in~\cite[Example~13.2]{Sandholm2015Handbook-of-gam} is a well-known example of a nonlinear contractive potential game. \\

\subsubsection{Linear Time Invariant Payoff Mechanism}
The payoff mechanism could be linear time invariant, as defined below. More generally, it could also be a payoff dynamical model (PDM) as thoroughly discussed in~\cite{Park2019From-Population}. \\

\begin{definition}{\bf [LTI]}\label{def:LTIPM} A payoff mechanism $\mathfrak{F}$ with input $\mathbf{x}$ and output $\mathbf{p}$ satisfies Assumption~\ref{ass:PayoffMech} and is of the linear time invatiant (LTI) type if it has a proper rational $n \times n$ transfer function matrix $F(s)$ whose poles have negative real part.  The associated stationary game is $\mathcal{F}_\mathfrak{F}(x)=F(0)x$.
\end{definition}

\section{Positive Correlation And Hybrid Rules}
\label{sec:NSandPC}

We proceed by defining a correlation function to establish the conditions for a rule to exhibit a property known as positive correlation. In \S\ref{subsec:telegen}, we motivate the relevance of positive correlation, arguing that most rules possess this property. In \S\ref{subsec:Canonicalrules}, we will define three widely studied canonical rule classes known to satisfy positive correlation. In \S\ref{subsec:HybridRules}, we will describe a general class of hybrid rules forming a cone spanned by the canonical classes. We will show that all hybrid rules within this cone are not only positively correlated but also satisfy another condition known as Nash stationarity. This characterization is crucial because our main result in \S\ref{sec:MainResult} applies to any rule that is positive correlated and Nash stationary. \\

\begin{definition} Given a rule $\mathcal{T}$, we define the correlation function $\wp:\mathbb{X} \times \mathbb{R}^n \to \mathbb{R}$ as follows:
\begin{equation} \label{eq:correlfuncdef}
\wp(x,p) : = p' \mathcal{V}(x,p), \quad x \in \mathbb{X}, \ p \in \mathbb{R}^n.
\end{equation} We say that $\mathcal{T}$ satisfies the \uline{positive correlation} property if the following condition holds for all $x$ in $\mathbb{X}$ and $p$ in $\mathbb{R}^n$:
\begin{equation}
\tag{PC}
\wp(x,p) \geq 0, \text{ and } \ \big (\wp(x,p)=0  \Leftrightarrow \mathcal{V}(x,p)=0 \big ).
\end{equation} 
\end{definition}
\vspace{.1 in}

Positive correlation requires that, away from an equilibrium ($\mathcal{V}(x,p) \neq 0)$, $\dot{x}$ forms an acute angle with $p$, i.e., ${p'\dot{x}>0}$.

\subsection{Tellegen And Universality Of Positive Correlation}
\label{subsec:telegen}

Inspired by~\cite{Tellegen1952A-general-netwo}, we derive from~(EDMb) and~(\ref{eq:correlfuncdef})
\begin{equation} \label{eq:Tellegen} \wp(x,p) = \frac{1}{2} \sum_{j=1}^{n} \sum_{i=1}^n  \tilde{U}_{ij} \tilde{I}_{ij},
\end{equation} where $\tilde{U}_{ij} : = p_j-p_i$ and $\tilde{I}_{ij}:=x_i \mathcal{T}_{ij}(x,p) - x_j \mathcal{T}_{ji}(x,p)$ is the net flow of agents switching from strategy $i$ to $j$. A simple way to derive the equality in~(\ref{eq:Tellegen}) is to expand the right hand side to obtain the left hand side via algebraic manipulation.

If we interpreted distinct strategies $i$ and $j$ as nodes of an electric network, agents as charged particles, and $(\tilde{U}_{ij},\tilde{I}_{ij})$ as the $(voltage,current)$ pair across the link from $i$ to $j$ then Kirchhoff's current law would have been equivalent to $\mathcal{V}(x,p)=0$, which using~(\ref{eq:correlfuncdef}) and~(\ref{eq:Tellegen}) would have led to Tellegen's Theorem~\cite{Tellegen1952A-general-netwo} expressed as ${\sum_{i=1}^{n} \sum_{j=1}^n  \tilde{U}_{ij} \tilde{I}_{ij}=0}$. \\

\begin{remark} \label{rem:telegen} It follows from~(\ref{eq:Tellegen}) that for $\mathcal{T}$ to satisfy (PC) it suffices that $x_i \mathcal{T}_{ij}(x,p) > x_j \mathcal{T}_{ji}(x,p) \implies p_j>p_i$, that is, more agents switch from $i$ to $j$ than from $j$ to $i$ only when~$p_j>p_i$. Hence, rational rules ought to satisfy (PC).
\end{remark}
\vspace{.1 in}

\subsection{Canonical Rule Classes Satisfying (PC)}
\label{subsec:Canonicalrules}

Remark~\ref{rem:telegen} is used in \cite{Sandholm2010Pairwise-compar,Sandholm2005Excess-payoff-d,Sandholm2010Population-Game} to prove that the canonical rule classes known as \uline{pairwise comparison}, \uline{excess payoff} and \uline{imitation} satisfy (PC). According to \cite[\S13.5]{Sandholm2015Handbook-of-gam}, these classes model a broad range of boundedly rational behaviors and \cite[\S2.3]{Sandholm2010Pairwise-compar} addresses their information requirements. We define these canonical classes below. \\

\begin{definition}{\bf Class $\mathbb{T}^\mathrm{CO}$:} \label{def:COProtocol} Any rule is said to be of the pairwise {\it comparison} (CO) type~\cite{Sandholm2010Pairwise-compar} if there is a locally Lipschitz map ${\phi:\mathbb{R}^n \to \mathbb{R}_{\geq 0}^{n \times n}}$ with which $\mathcal{T}$ can be written~as
\begin{equation}
\label{eq:defC}
\mathcal{T}_{ij}(x,p) \underset{\mathrm{CO}}{=} \phi_{ij}(p), \ x \in \mathbb{X}, \ p \in \mathbb{R}^n,
\end{equation} where $\phi_{ij}(p) > 0$ when $p_{j}>p_i$ and $\phi_{ij}(p) = 0$ when $p_j \leq p_i$. We use $\mathbb{T}^\mathrm{CO}$ to denote the set of all CO rules.
\end{definition}

\begin{example}\label{example:Smith} The well-known Smith's rule~\cite{Smith1984The-stability-o} $\mathcal{T}^{\text{\tiny Smith}}$ is in $\mathbb{T}^\mathrm{CO}$ and can be specified by selecting ${\phi_{ij}^{\text{\tiny Smith}}(p) :=  [p_j-p_i]_+}$. 
\end{example}
\vspace{.1 in}

Before we proceed to define the class of excess payoff rules, we define the so-called \uline{excess payoff vector} $\hat{p}$ as $${\hat{p}_i:=p_i-p'x}, \quad i \in \{1,\ldots,n\},$$ where the population's average is $p'x$ and the set of possible excess payoffs is defined as $\mathbb{R}^n_*:=\{\hat{p} \ | \ p \in \mathbb{R}^n, \ x \in \mathbb{X} \}$. \\

\begin{definition}{\bf Class $\mathbb{T}^\mathrm{EP}$:} \label{def:EP} Any protocol is said to be of the {\it excess payoff} (EP) type~\cite{Sandholm2005Excess-payoff-d} if there is a locally Lipschitz map ${\varphi:\mathbb{R}_*^n \to \mathbb{R}^n }$, such that $\mathcal{T}$ can be recast as
\begin{equation}
\label{eq:defSEPT}
\mathcal{T}_{ij}(x,p) \underset{\mathrm{EP}}{=} \varphi_{j}(\hat{p}), \quad \hat{p}_{j}:= p_j-\sum_{i=1}^n x_i p_i.
\end{equation} A valid choice of $\varphi$ must satisfy $\hat{p}'\varphi(\hat{p}) > 0$ for every ${\hat{p} \in \mathrm{int}(\mathbb{R}_*^n)}$. We use $\mathbb{T}^\mathrm{EP}$ to denote the set of all EP rules. 
\end{definition}

\begin{example} \label{ex:BNN} The classic Brown-von Neumann-Nash (BNN) rule~\cite{Brown1950Solutions-of-ga} is in $\mathbb{T}^\mathrm{EP}$ and is specified by $\varphi_{j}^{\text{\tiny BNN}}(\hat{p}) := [\hat{p}_{j}]_+$.
\end{example}
\vspace{.1 in}
\begin{example}{\bf [Approximate best response]} \label{ex:ApprxBestResponse} After a small modification of~\cite[\S2.4.2]{Sandholm2005Excess-payoff-d} to guarantee Lipschitz continuity, we obtain an approximate best response (ABR) rule in $\mathbb{T}^\mathrm{EP}$ denoted as $\mathcal{T}^{\text{\tiny ABR}}$ and specified using~(\ref{eq:defSEPT}) with
\begin{equation} \varphi_{j}^{\text{\tiny ABR}}(\hat{p}) := \frac{[\hat{p}_{j}]_+^k}{\sum_{i=1}^n [\hat{p}_{i}]_+^k + \epsilon^k},
\end{equation} where $\epsilon \in (0,1)$ and $k$ is a positive integer. The larger $k$ is the closer $\mathcal{T}^{\text{\tiny ABR}}$ is to the best response rule.
\end{example}
\vspace{.1 in}

\begin{definition}{\bf Class $\mathbb{T}^\mathrm{I}$:} \label{def:ImitativeProtocol} Any rule is said to be of the {\it imitative} (I) class if there is a locally Lipschitz map ${\psi:\mathbb{X} \times \mathbb{R}^n \to \mathbb{R}_{\geq 0}^{n \times n}}$ such that $\mathcal{T}$ can be written as
\begin{equation}
\label{eq:defImitation}
\mathcal{T}_{ij}(x,p) \underset{\mathrm{I}}{=} x_j \psi_{ij}(x,p),
\end{equation} where $\psi$ satisfies the following monotonicity condition for all $i$, $j$ and $k$ in $\{1,\ldots,n\}$:
\begin{equation}
p_j \geq p_i \Leftrightarrow \psi_{kj}(x,p)-\psi_{jk}(x,p) \geq  \psi_{ki}(x,p)-\psi_{ik}(x,p).
\end{equation}
 We use $\mathbb{T}^\mathrm{I}$ to denote the set of all imitative rules.
\end{definition}

\begin{example}\label{ex:replicator} The so-called \uline{replicator} rule~$\mathcal{T}^\mathrm{R}$ is a ubiquitous example in $\mathbb{T}^\mathrm{I}$ obtained as $\psi^{\text{\tiny R}}_{ij}(x,p) =[p_j-p_i]_+$. 
\end{example}
\vspace{.1 in}
\begin{remark}\label{rem:NoBestResponse} As we discussed in \S\ref{subsec:ContribAndLim}, we excluded the so-called `best response' rule~\cite[13.5.2]{Sandholm2015Handbook-of-gam} from our analysis to avoid having to use Carathéodory solutions for differential inclusions and a modified (PC) notion. Including it wouldn't have significantly changed our results but would have complicated the paper and detracted from our focus on using CCW to study stability. Fortunately, Example~\ref{ex:ApprxBestResponse} shows that we can approximate the best response rule to an arbitrary degree of accuracy by a rule in $\mathbb{T}^\mathrm{EP}$.
\end{remark}

\subsection{Convex Cones, Hybrid Rules and Key Properties}
\label{subsec:HybridRules}

The classes $\mathbb{T}^\mathrm{CO}$, $\mathbb{T}^\mathrm{EP}$ and $\mathbb{T}^\mathrm{I}$ represent a range of strategic behaviors and preferences of agents. Specifically, rules in $\mathbb{T}^\mathrm{CO}$ allow agents to adopt any strategy that improves their current payoffs, while rules in $\mathbb{T}^\mathrm{EP}$ permit changes only to strategies exceeding the current average population payoff. To account for imitation of ``popular" strategies, rules in $\mathbb{T}^\mathrm{I}$ incorporate a multiplicative term $x_i$. These learning classes are broad because they are convex cones, meaning they include any positive linear combination of rules within them. However, it is unrealistic to expect strict adherence to a single class of rules. As \cite[p.~5]{Sandholm2010Pairwise-compar} suggests, agents are likely to follow hybrid rules as defined below combining elements from multiple classes. \\

\begin{definition}{\bf (\bf Hybrid rule)} \label{def:HybLearnRule} We qualify a rule $\mathcal{T}$ as hybrid if it is expressible as 
\begin{multline}
\label{eq:HybLearnRule}
\mathcal{T} = \alpha^\mathrm{I} \mathcal{T}^\mathrm{I} + \alpha^\mathrm{CO} \mathcal{T}^\mathrm{CO}+\alpha^\mathrm{EP} \mathcal{T}^\mathrm{EP}+\tilde{\alpha}\tilde{\mathcal{T}}, \\ \alpha^\mathrm{CO} + \alpha^\mathrm{EP} > 0,
\end{multline} where $\alpha^\mathrm{I}$, $\alpha^\mathrm{CO}$, $\alpha^\mathrm{EP}$ and $\tilde{\alpha}$ are nonnegative weights, with ${\mathcal{T}^\mathrm{I} \in \mathbb{T}^\mathrm{I}}$, $\mathcal{T}^\mathrm{CO} \in \mathbb{T}^\mathrm{CO}$, $\mathcal{T}^\mathrm{EP} \in \mathbb{T}^\mathrm{EP}$ and \uline{$\tilde{\mathcal{T}}$ is any rule satisfying (PC)}. We include $\tilde{\mathcal{T}}$ to allow for the case when $\mathcal{T}$ cannot be exactly expressed as a conic combination of rules in the canonical classes.
 \end{definition}

A hybrid rule $\mathcal{T}$ can be rewritten as:
\begin{equation}
\label{eq:InterpHybLearnRule}
\mathcal{T} = \bar{\alpha} \left ( \frac{\alpha^\mathrm{I} }{\bar{\alpha}} \mathcal{T}^\mathrm{I} + \frac{\alpha^\mathrm{CO} }{\bar{\alpha} } \mathcal{T}^\mathrm{CO}+ \frac{\alpha^\mathrm{EP}}{\bar{\alpha} } \mathcal{T}^\mathrm{EP} +\frac{\tilde{\alpha}}{\bar{\alpha} } \tilde{\mathcal{T}} \right ),
\end{equation} where $\bar{\alpha} := \alpha^\mathrm{I} + \alpha^\mathrm{CO} + \alpha^\mathrm{EP}+\tilde{\alpha}$ acts as a multiplicative factor. \\

Notice that the constraint $\alpha^\mathrm{CO} + \alpha^\mathrm{EP} > 0$ prevents hybrid rules from being purely imitative, which is a technical condition to avoid the unrealistic so-called extinction property according to which a strategy that is currently not adopted could never be adopted in the future.

There is a probabilistic interpretation\footnote{See~\cite[\S4]{Sandholm2003Evolution-and-e},\cite[\S IV]{Park2019From-Population} and~\cite[\S2]{Kara_2023aa} as well as references therein.} for how a hybrid rule models the behavior of a strategic agent. Namely, according to \cite{Sandholm2010Pairwise-compar}, (\ref{eq:InterpHybLearnRule}) can be interpreted as modeling an agent that at every strategy revision opportunity selects a imitative, CO or EP rule with probabilities $\frac{\alpha^\mathrm{I}}{\bar{\alpha}}$, $\frac{\alpha^\mathrm{CO}}{\bar{\alpha}}$ and $\frac{\alpha^\mathrm{EP}}{\bar{\alpha}}$, respectively. With probability $\frac{\tilde{\alpha}}{\bar{\alpha}}$, the agent could follow any rule satisfying~(PC) (see Remark~\ref{rem:telegen}). \\

\begin{proposition} \label{prop:HybridPCNS} If $\mathcal{T}$ is a hybrid rule expressible as~(\ref{eq:HybLearnRule}), then $\mathcal{T}$ satisfies (PC) with the correlation function
\begin{equation}
\label{eq:HybridCorrelationfunction}
\wp = \alpha^\mathrm{I} \wp^\mathrm{I} + \alpha^\mathrm{CO} \wp^\mathrm{CO}+\alpha^\mathrm{EP} \wp^\mathrm{EP}+\tilde{\alpha}\tilde{\wp},
\end{equation} where $\wp^\mathrm{I}$,  $\wp^\mathrm{CO}$, $\wp^\mathrm{EP}$ and $\tilde{\wp}$ are, respectively, the correlation functions for $\mathcal{T}^\mathrm{I}$,  $\mathcal{T}^\mathrm{CO}$, $\mathcal{T}^\mathrm{EP}$ and $\tilde{\mathcal{T}}$. In addition, $\wp$ satisfies the equivalence
\begin{equation}
\label{eq:NashStationarity}
\wp(x,p) = 0 \Leftrightarrow x \in \mathfrak{B}(p), \quad x \in \mathbb{X}, \ p \in \mathbb{R}^n,
\end{equation} where $\mathfrak{B}:\mathbb{R}^n \to 2^\mathbb{X}$ is the best response map defined as:
\begin{equation*}
\mathfrak{B}(p) := \Big \{z \in \mathbb{X} \ \Big | \ p'z = \max_{1\leq i \leq n} p_i \Big \}.
\end{equation*}
\end{proposition}
\vspace{.1 in}

\begin{remark}{\bf [Nash Stationarity (NS)]} \label{rem:(NS)} Subject to $\mathcal{T}$ satisfying (PC), the condition~(\ref{eq:NashStationarity}) is equivalent to the so-called Nash stationarity (NS) property according to which $\mathcal{V}(x,p)=0$ if and only if $x \in \mathfrak{B}(p)$, that is, $\mathcal{V}(x,p)=0$ if and only if $x$ is a best response to $p$. We will use (NS) in the proof of Proposition~\ref{prop:HybridPCNS} and in proving our main result in~\S\ref{sec:MainResult}. Rules satisfying both (NS) and (PC) are also called {\it well-behaved}~\cite{Sandholm2005Excess-payoff-d}.
\end{remark}
\vspace{.1 in}
\paragraph*{\bf Proof of Proposition~\ref{prop:HybridPCNS}} We note that~\cite[Theorem~5.4.9]{Sandholm2010Population-Game}, \cite[Theorem~1]{Sandholm2010Pairwise-compar} and \cite[Theorem~3.1]{Sandholm2005Excess-payoff-d} state, respectively, that $\mathcal{T}^\mathrm{I}$,  $\mathcal{T}^\mathrm{CO}$ and $\mathcal{T}^\mathrm{EP}$ satisfy (PC). These theorems and (\ref{eq:HybridCorrelationfunction}), which we obtain by linearity, imply that $\mathcal{T}$ satisfies (PC).  We can use Remark~\ref{rem:(NS)} and \cite[Theorem~2]{Sandholm2010Pairwise-compar} to prove~(\ref{eq:NashStationarity}). $\square$
\vspace{.1 in}

\begin{example}\label{ex:HybridRules} Examples of hybrid rules as in Definition~\ref{def:HybLearnRule}, include $\mathcal{T}^\mathrm{a}$, $\mathcal{T}^\mathrm{b}$, $\mathcal{T}^\mathrm{c}$ and $\mathcal{T}^\mathrm{d}$ as follows
\begin{itemize}
\item[] $\mathcal{T}_{ij}^\mathrm{a}(x,p) = 2x_j[p_j-p_i]_+^2+3j(e^{[p_j-p_i]_+}-1) +4 [\hat{p}_j]_+^2$,
\item[]
\item[] $\mathcal{T}_{ij}^\mathrm{b}(x,p) = \underbrace{x_j[p_j-p_i]_+}_{\mathcal{T}^\mathrm{R}}+ 0.01 \underbrace{[p_j-p_i]_+}_{\mathcal{T}^{\text{\tiny Smith}}} $,
\item[]
\item[] $\mathcal{T}_{ij}^\mathrm{c}(x,p) = 0.01 \underbrace{[p_j-p_i]_+}_{\mathcal{T}^{\text{\tiny Smith}}} +  \underbrace{\frac{[\hat{p}_j]_+^5}{\sum_{i=1}^n [\hat{p}_i]_+^5 +10^{-5}}}_{\mathcal{T}^{\text{\tiny ABR}}}$,
\item[]
\item[] $\mathcal{T}_{ij}^\mathrm{d}(x,p) = 0.2 \mathcal{T}_{ij}^\mathrm{a}(x,p) + 3 \mathcal{T}_{ij}^\mathrm{b}(x,p) + 40\mathcal{T}_{ij}^\mathrm{c}(x,p)$,
\end{itemize} where $\mathcal{T}^\mathrm{b}$ and $\mathcal{T}^\mathrm{c}$ can be viewed as approximations of the replicator rule $\mathcal{T}^\mathrm{R}$ in Example~\ref{ex:replicator} and the approximate best response rule $\mathcal{T}^{\text{\tiny ABR}}$ in Example~\ref{ex:ApprxBestResponse}, respectively.
\end{example} 
\vspace{.1 in}

\section{Counterclockwise Dissipativity And \\ Main Results}
\label{sec:MainResult}
We start by adapting to our framework the concept of counterclockwise systems proposed in~\cite{Angeli2006Systems-With-Co} to study stability and robustness of nonlinear feedback systems. \\

\begin{definition}{\bf [CCW]} A payoff mechanism $\mathfrak{F}$ is counterclockwise dissipative (or CCW for short), if 
\begin{equation} \label{eq:CCWDEf} \alpha_\mathfrak{F} : = - \inf_{T>0,\mathbf{x} \in \mathscr{X}} \int_0^T \dot{p}'(t)x(t)dt < \infty.
\end{equation}
\end{definition} 

\subsection{CCW Payoff Mechanisms And Potential Games}

The following theorem states that potential games are the only CCW memoryless payoff mechanisms (see Definition~\ref{def:game}).
\vspace{.2 in}
\begin{theorem}{}{CCWGameIsPotential} 
    A game is potential if and only if it is CCW.
\end{theorem}
\vspace{.2 in}
\paragraph*{\bf Proof of Theorem~\ref{thm:CCWGameIsPotential}} Let $\mathcal{F}: x(t) \mapsto p(t)$ specify a game. To prove $(\mathcal{F}\text{ is potential}) \implies (\mathcal{F}\text{ is CCW})$, we use~(\ref{eq:PotentialGame}) and integration by parts to write for any $\mathbf{x}$ in $\mathscr{X}$, and $T>0$
\begin{equation*}
- \int_0^T \dot{p}'(t)x(t)dt  \leq 2 \big ( \max_{x \in \mathbb{X}} \|\mathcal{F}(x)\|_\infty + \max_{x \in \mathbb{X}} f(x)\big)<\infty, 
\end{equation*} where we use continuity of $\mathcal{F}$ and $f$, and compactness of $\mathbb{X}$ to establish boundedness. 

We now proceed to show that $(\mathcal{F}\text{ is not potential}) \implies (\mathcal{F}\text{ is not CCW})$. If $\mathcal{F}$ is not potential then it is not a conservative vector field in $\mathbb{X}$ and there is $\mu>0$ and a periodic $\mathbf{x}$ in $\mathscr{X}$, with $x(0)=x(kT)$ for any natural $k$, such that
$\int_0^T p'(t)\dot{x}(t)dt = \mu$. Using integration by parts and periodicity of $\mathbf{x}$ we obtain$ 
\int_0^{kT} \dot{p}'(t)x(t)dt = - k \mu$ disproving CCW. $\square$

\subsection{CCW Convex Cone and Negative Imaginary Systems}

Modifying the proof of \cite[Theorem~1]{Angeli2006Systems-With-Co} one can show that an LTI payoff mechanism (Definition~\ref{def:LTIPM}) is CCW if the negative imaginary (NI) condition $j(F(j\omega)-F'(-j\omega))\succeq 0$ holds for all $\omega>0$. The concept of (NI) for multiple inputs and outputs was proposed in~\cite{Lanzon2008Stability-Robus} to study stability and robustness. Relevant subclasses of (NI) systems and a dissipative framework for stability are studied in~\cite{Lanzon2023Characterizatio}. \\

\begin{remark} \label{rem_conepayoffs} If $\mathfrak{F}$ and $\mathfrak{G}$ are CCW then $\xi \mathfrak{F}+\zeta \mathfrak{G}$ is also CCW for any non-negative $\xi$ and $\zeta$, implying that the set of CCW payoff mechanisms is a convex cone (see also \cite[Proposition~III.1]{Angeli2006Systems-With-Co}). As a result, the already large class of (NI) systems~\cite{Petersen2016Negative-imagin} and nonlinear generalizations~\cite{Ghallab2018Extending-Negat} that are CCW can be combined with potential games (see Theorem~\ref{thm:CCWGameIsPotential}) to form a broad convex cone of CCW payoff mechanisms. 
\end{remark}

The following is an example interpretable as a potential game with an LTI additive perturbation we denote as $\mathfrak{G}$.  

\begin{example}\label{ex_NIAndPotential} For a potential game $\mathcal{F}$, a given time-constant $\lambda^{-1}>0$, $b$ in $\mathbb{R}^n$, and $A=A'$ in $\mathbb{R}^{n \times n}$, consider
\begin{subequations}
\label{eq_examplepayoff}
\begin{align}
\dot{q}(t) &= \lambda \big ( Ax(t) + b - q(t) \big ), \qquad t\geq0, \quad q(0)=0, \\
p(t) &= \mathcal{F}\big ( x(t) \big ) + k \lambda \big ( A x(t) + b - q(t) \big)
\end{align}
\end{subequations} where $k$ is a real constant satisfying $k A \preceq 0$.
\end{example}

The transfer function matrix $G(s)$ from $\mathbf{x}$ to $k \lambda ( A \mathbf{x} - \mathbf{q})$ is NI satisfying ${j(G(j\omega)-G'(-j\omega)) = -2 k \lambda^2 \tfrac{\omega}{\omega^2+\lambda^2}A}$, and we conclude from Remark~\ref{rem_conepayoffs} that~(\ref{eq_examplepayoff}) is CCW with stationary game $\mathcal{F}(x)$. Notice that the effect of $b$ in~(\ref{eq_examplepayoff}) vanishes exponentially fast and plays no role in testing for CCW. 

\begin{remark}\label{rem:detailsaboutexampls} Example~\ref{ex_NIAndPotential} generalizes well-motivated examples in~\cite{Fox2013Population-Game}. Namely,\cite[(83)]{Fox2013Population-Game} follows by specializing~(\ref{eq_examplepayoff}) with $k=1$ and $\mathcal{F} ( x )=Ax+b$, where $A \prec 0$. The example \cite[(77)]{Fox2013Population-Game} follows by specializing~(\ref{eq_examplepayoff}) with $k=-\lambda^{-2}$, and $\mathcal{F} ( x )=\lambda^{-1}(Ax+b)$. With this parameter choice, we must impose $A \succeq 0$, such as in a coordination game, in contrast with~\cite[(77)]{Fox2013Population-Game} where $A \prec 0$.
\end{remark}

\subsection{Evolutionary Nash Equilibrium Learning Theorem}
\label{subsec:StabThm}

The feedback interconnection of two strict CCW systems has important convergence properties~\cite{Angeli2006Systems-With-Co}. If a rule $\mathcal{T}$ satisfies (PC), then $\int_0^T \dot{x}'(t)p(t) , dt \geq 0$ holds for any $\mathbf{p} \in \mathscr{P}$ and ${T > 0}$, i.e., the (EDM) is itself CCW. This fact and~\cite{Angeli2006Systems-With-Co} motivated our investigation into the convergence properties of $\mathbf{x}$ when $\mathcal{T}$ satisfies (PC) and the payoff mechanism~$\mathfrak{F}$ is CCW.

The following theorem states that imposing (PC) and~(\ref{eq:NashStationarity}) on $\mathcal{T}$, combined with a $\mathfrak{F}$ that is CCW, ensures convergence of $x(t)$ to the Nash equilibrium set of $\mathcal{F}_\mathfrak{F}$ (see Assumption~\ref{ass:PayoffMech}.2). Unlike in \cite{Angeli2006Systems-With-Co}, the Nash equilibrium concept is central to our theorem and its proof must address $\mathfrak{F}$ not being strictly~CCW.

\vspace{.2 in}
\begin{theorem}{}{Main}
    Consider that $\mathcal{T}$ is a rule satisfying (PC) and~(\ref{eq:NashStationarity}), or, equivalently, suppose that $\mathcal{T}$ satisfies (PC) and (NS) (See Remark~\ref{rem:(NS)}). If  the payoff mechanism $\mathfrak{F}$ is CCW, then, for the feedback system's solution $\mathbf{x}$, it holds that
\begin{equation} \label{eq:MainThm}
\lim_{t \rightarrow \infty} \|\dot{x}(t)\| \underset{(a)}{=} 0 \text{ and } \lim_{t \rightarrow \infty} \inf_{y \in \mathbb{NE}(\mathcal{F}_\mathfrak{F})} \|x(t) - y \| \underset{(b)}{=} 0
\end{equation} for any $x(0)$ in $\mathbb{X}$, where $\mathbb{NE}(\mathcal{F}_\mathfrak{F})$ is the Nash equilibria set\footnote{In our context, Nash equilibria should be interpreted in the mass-action sense explained in~\cite{Weibull1995The-mass-action,Jr.1951Non-Cooperative}.}
\begin{equation*}
\mathbb{NE}(\mathcal{F}_\mathfrak{F}) : = \Big \{ x \in \mathbb{X} \ \big | \ x \in \mathfrak{B} \big (\mathcal{F}_\mathfrak{F}(x) \big ) \Big \}.
\end{equation*}
\end{theorem}
\vspace{.2 in}
 Hence, \uline{evolutionary Nash equilibrium learning}, as expressed in~(\ref{eq:MainThm}), is guaranteed when the conditions of Theorem~\ref{thm:Main} hold. In addition, we infer the following facts.

\begin{itemize}
    \item[\textbf{(i)}] Using $(a)$ in (\ref{eq:MainThm}) and Assumption~\ref{ass:PayoffMech}.2, we conclude that for sufficiently large $t$, $\mathfrak{F}$ behaves approximately as a stationary game: $p(t) \simeq \mathcal{F}_{\mathfrak{F}}(x(t))$. The continuity of $\mathcal{F}_{\mathfrak{F}}$ implies that any accumulation point $(x^*,p^*)$ of $(\mathbf{x},\mathbf{p})$ satisfies $p^* = \mathcal{F}_{\mathfrak{F}}(x^*)$, and the continuity of $\mathcal{V}$ ensures $\mathcal{V}(x^*,p^*) = 0$. Consequently, all accumulation points are equilibria and there are \underline{no nonconstant limit cycles}.
    \item[\textbf{(ii)}] According to $(b)$ in (\ref{eq:MainThm}), $\mathbb{NE}(\mathcal{F}_{\mathfrak{F}})$ predicts the long-term evolution of $\mathbf{x}$. By the definition of  $\mathbb{NE}(\mathcal{F}_{\mathfrak{F}})$, (i) and $(b)$ in (\ref{eq:MainThm}) imply for large $t$ that $x(t)$ tends to $\mathfrak{B}(p(t))$, i.e., $x(t)$ tends to become a best response to~$p(t)$.
    \item[\textbf{(iii)}] Provided that (PC) and (NS) are satisfied, the theorem does not require any specific knowledge about $\mathcal{T}$ and imposes no constraints on coordination among the agents\footnote{Example~\ref{example:Smith} specifies a fully decentralized rule satisfying (PC) and (NS).}.
\end{itemize}

\paragraph*{\bf Proof of Theorem~\ref{thm:Main}}
Let $x(0)$ in $\mathbb{X}$ be an initial condition and $(\mathbf{x},\mathbf{p})$ be the associated solution trajectory pair. Using integration by parts, Assumption~\ref{ass:PayoffMech}.3, (\ref{eq:correlfuncdef}) and (\ref{eq:CCWDEf}), we obtain
\begin{equation} \label{eq:IntegralCorrBounded}
\int_0^T \wp \big (x(t),p(t))dt \leq 2 \beta_\mathfrak{F} + \alpha_\mathfrak{F}, \quad T>0.
\end{equation} Since $\wp$ is a Lipschitz continuous map and $(x,p)(t)$ is Lipschitz continuous w.r.t time, we conclude that $\wp \big (x(t),p(t))$ is also Lipschitz continuous w.r.t. time. From this fact and~(\ref{eq:IntegralCorrBounded}) we can use {B}arb\u{a}lat's Lemma~\cite{Farkas2016Variations-on-B} to conclude that $\lim_{t \rightarrow \infty} \wp \big (x(t),p(t)) = 0$, which also implies $\lim_{t \rightarrow \infty} \mathcal{V} \big (x(t),p(t)) = 0$ leading to $(a)$ in~(\ref{eq:MainThm}). This implication results from (PC) and the facts that $\mathcal{V}$ and $\wp$ are continuous, and $(\mathbf{x},\mathbf{p})$ takes values in the bounded set $\mathbb{X} \times [-\beta_\mathfrak{F},\beta_\mathfrak{F}]^n$, where $\beta_\mathfrak{F}$ comes from Assumption~\ref{ass:PayoffMech}.3.

We now proceed to prove $(b)$ in~(\ref{eq:MainThm}) by contradiction. To do so we will use $(a)$ in~(\ref{eq:MainThm}) and assume that $(b)$ in~(\ref{eq:MainThm}) did not hold. If this were the case, $(\mathbf{x},\mathbf{p})$ would have had an accumulation point $(x^*,p^*)$ satisfying $\inf_{y \in \mathbb{NE}(\mathcal{F}_\mathfrak{F})} \|x^* - y \| >0$. Furthermore, $(a)$ in~(\ref{eq:MainThm}) and Assumption~\ref{ass:PayoffMech}.2 we would have implied that ${p^*=\mathcal{F}_\mathfrak{F}(x^*)}$. Consequently, we would have been able to infer $x^* \notin \mathbb{NE}(\mathcal{F}_\mathfrak{F})$ and $x^* \notin \mathfrak{B}(p^*)$, but from~(\ref{eq:NashStationarity}) that would have implied  $\wp(x^*,p^*) >0$, which would have contradicted $\lim_{t \rightarrow \infty} \wp \big (x(t),p(t)) = 0$ that we already proved above. $\square$ \\

By allowing for dynamic $\mathfrak{F}$, we extend the original framework of memoryless payoff mechanisms in \cite{Sandholm2001Potential-games}, where the focus is in potential games. Specifically, using Theorem~\ref{thm:CCWGameIsPotential}, the result presented in \cite[Theorem 4.5(ii)]{Sandholm2001Potential-games} for potential games can be derived as a particular case of Theorem~\ref{thm:Main} applied to the specific case of a memoryless payoff mechanism.
 \\

Theorem~\ref{thm:Main}, combined with Proposition~\ref{prop:HybridPCNS}, directly leads to the following corollary, which states the \uline{primary conclusion of Theorem~\ref{thm:Main} we are seeking to obtain in this paper}. Namely, with a CCW payoff mechanism $\mathfrak{F}$, the corollary guarantees evolutionary Nash equilibria learning for any hybrid rule (Definition~\ref{def:HybLearnRule}). This result is notably robust, as it applies regardless of the specific hybrid rule used, provided that the rule is known to be hybrid.

\vspace{.2 in}
\begin{corollary}{}{} If $\mathcal{T}$ is a hybrid rule (see Definition~\ref{def:HybLearnRule}) and  the payoff mechanism $\mathfrak{F}$ is CCW then (\ref{eq:MainThm}) holds for the feedback system's solution $\mathbf{x}$ for any $x(0)$ in $\mathbb{X}$.
\end{corollary}
\vspace{.2 in}

\subsection{Comparing With Other Passivity Approaches}
\label{subsec:ComparisonPassiveApproaches}

A system theoretic passivity approach valid for $\mathbb{T}^\mathrm{CO}$ and $\mathbb{T}^\mathrm{EP}$ was first introduced in \cite{Fox2013Population-Game}, where $\delta$-passivity generalizes contractive games to allow for dynamic payoffs. The convergence results in \cite{Fox2013Population-Game} require both the (EDM) and $-\mathfrak{F}$ to be $\delta$-passive. Since \cite{Park2018Passivity-and-e} showed that the (EDM) for the replicator rule $\mathcal{T}^{\mathrm{R}}$ (Example~\ref{ex:replicator}) is not $\delta$-passive, we conjecture that $\delta$-passivity will not hold for any hybrid rule (\ref{eq:HybLearnRule}) with $\alpha^{\mathrm{I}}>0$.

The work in~\cite{Mabrok2021Passivity-Analy} established that {\bf (i)} the (EDM) associated with $\mathcal{T}^{\mathrm{R}}$ is equilibrium independent passive (EIP)~\cite{Hines2011Equilibrium-ind}, ensuring convergence to equilibria when $-\mathfrak{F}$ is strictly passive, and {\bf (ii)} adding integral or lead-lag second-order action to the payoff makes the resulting higher-order (EDM) NI (or CCW), ensuring stability for any strictly NI $\mathfrak{F}$. EIP was also used to show convergence for a modified $\mathcal{T}^{\mathrm{R}}$ with exponentially-discounted learning~\cite{Gao2020On-Passivity-Re}. In our work, $\mathcal{T}^{\mathrm{R}}$ cannot be considered in isolation but only ``approximately" because it is not in the cone of hybrid rules~(\ref{eq:HybLearnRule}) due to the constraint $\alpha^\mathrm{CO} + \alpha^\mathrm{EP} > 0$ needed for Nash stationarity~(\ref{eq:NashStationarity}). Conversely, the approach in~\cite{Mabrok2021Passivity-Analy} is not applicable to our framework because EIP is not known to hold for the hybrid rules~(\ref{eq:HybLearnRule}). Unlike (ii), our work does not consider higher-order (EDM), so no direct comparison can be made, except noting that NI and CCW concepts are used in both.

Hence, existing passivity-based approaches related to $\mathcal{T}^{\mathrm{R}}$ are not, in general, applicable to rules in $\mathbb{T}^{\mathrm{CO}}$ and $\mathbb{T}^{\mathrm{EP}}$, and vice versa. Our approach is the first to support hybrid rules incorporating elements of $\mathbb{T}^{\mathrm{CO}}$, $\mathbb{T}^{\mathrm{EP}}$, and $\mathbb{T}^\mathrm{I}$, which includes $\mathcal{T}^{\mathrm{R}}$ as a particular case. Notice that~\cite{Certorio_2024aa} studies $\delta$-passivity for hybrid rules that include the best response rule, which we can approximate with arbitrary accuracy as in Example~\ref{ex:ApprxBestResponse}. Although the work~\cite{Certorio_2024aa} is the first to study $\delta$-passivity for hybrid rules, it still has major gaps in comparison to the hybrid rules we consider here. Namely, the approach in~\cite{Certorio_2024aa} {\bf (i)} does not consider $\mathbb{T}^\mathrm{I}$, {\bf (ii)} it cannot handle conic combinations between $\mathbb{T}^\mathrm{EP}$ and $\mathbb{T}^\mathrm{CO}$ beyond two strategies, and {\bf (iii)} it considers only the so-called separable subclass of $\mathbb{T}^\mathrm{EP}$ denoted as SEPT. Since the best response rule cannot be approximated by an SEPT rule, the best response had to be considered directly in~\cite{Certorio_2024aa} requiring complicated notions of solution for differential inclusions. \\

While our approach accommodates the broadest set of rules, it is best understood as complementary to existing passivity-based methods. Each approach has unique strengths, addressing cases that others may not. This becomes evident when we examine the requirements on the payoff mechanism. Our method requires $\mathfrak{F}$ to be CCW, unlike previous works that necessitate $\mathfrak{F}$ to be strictly NI, $-\mathfrak{F}$ to be strictly EIP~\cite{Mabrok2021Passivity-Analy}, or $-\mathfrak{F}$ to be $\delta$-passive~\cite{Fox2013Population-Game,Certorio_2024aa}. Generally, these requirements are not directly comparable. We illustrate this below:
\begin{itemize}
    \item $\mathcal{F}(x)=Ax$ is CCW for any $A$ in $\mathbb{R}^{n \times n}$ satisfying ${(w-x)'(A-A')(v-x)=0}$ for all $w,v,x \in \mathbb{X}$. This condition is satisfied for any symmetric $A$.
    \item $\mathcal{F}(x)=Ax$ is not strictly NI for any $A$ in $\mathbb{R}^{n \times n}$.
    \item $-\mathcal{F}(x)=-Ax$ is EIP and $\delta$-passive for any $A$ in $\mathbb{R}^{n \times n}$ satisfying $(v-x)'(A+A')(v-x) \geq 0$ for $v,x \in \mathbb{X}$. This includes any positive-semidefinite nonsymmetric $A$.
    \item Remark~\ref{rem:detailsaboutexampls} illustrates a case there $-\mathfrak{F}$ is $\delta$-passive and $\mathfrak{F}$ is CCW and another case where either $-\mathfrak{F}$ is $\delta$-passive or $\mathfrak{F}$ is CCW, but not both.
\end{itemize}

\section{Numerical Example}
\label{sec:Simulation}

To illustrate our results, we adopt a payoff mechanism based on Example~\ref{ex_NIAndPotential} for $3$ strategies ($n=3$), a potential game specified as
$\mathcal{F}_i(x) = 1-x_i$, and parameters $k = -1$, $\lambda = 5$,
\begin{align*}
    b &= \frac{1}{(k \lambda)}\begin{bmatrix}
        2 \\0 \\0
    \end{bmatrix} , &
    A &= \begin{bmatrix}
        0 &0 &0\\ 0 &1 &0\\ 0 &0 &1
    \end{bmatrix}.
\end{align*}
As described in Example~\ref{ex_NIAndPotential}, $A=A'\succeq 0$ and $k=-1$ makes the payoff dynamic model be CCW with stationary game $\mathcal{F}$. Due to the value of $b$ agents initially have a strong preference for strategy 1, but that effect vanishes exponentially fast.

\begin{figure}[hb]
    \center
    \scalebox{1.0}{\input{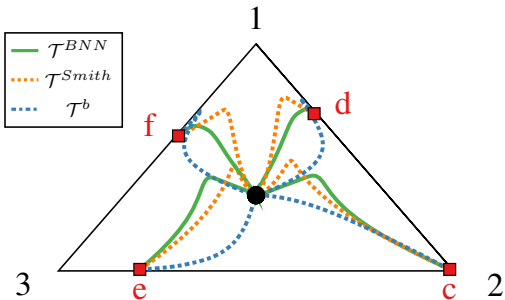}}
    \caption{Trajectories for the population state $x(t)$ converging to $\mathbb{NE}(\mathcal{F})$ under different initial conditions and learning rules, respectively, ${x(0)\in \{c,d,e,f\}}$ and $\mathcal{T}\in \{\mathcal{T}^{\mathrm{BNN}}, \mathcal{T}^{\mathrm{Smith}}, \mathcal{T}^{\mathrm{b}}\}$.}
    \label{fig:pdm_trajectories}
\end{figure}

We ran simulations \cite{jair_2024_12727062} for rules $\mathcal{T}^{\mathrm{BNN}}$ (Example~\ref{ex:BNN}), $\mathcal{T}^{\mathrm{Smith}}$~(Example~\ref{example:Smith}), and $\mathcal{T}^{\mathrm{b}}$ from Example~\ref{ex:HybridRules}. The initial conditions labelled as $\{ c,d,e,f \}$ are specified as
$c:= [0\; 1\; 0]'$, $d:= [0.7\;  0.3\;  0]'$, ${e:= [0\;  0.2\;  0.8]'}$, and ${f:= [0.6\;  0\;  0.4]'}$. The resulting trajectories can be seen in Fig.~\ref{fig:pdm_trajectories}, with the red squares and black circle denoting the initial and final points of the trajectories, respectively. As expected, all trajectories converge to the unique Nash equilibrium of the game $\mathcal{F}$, $\mathbb{NE}(\mathcal{F})=\{[0.\bar{3} \;0.\bar{3} \;0.\bar{3}]'\}$.

\section{Conclusions}
\label{sec:Conclusions}

In this paper, we addressed the problem of achieving evolutionary Nash equilibrium learning in large populations of agents engaged in strategic, non-cooperative interactions. Our main focus was to bridge the gap between $\delta$-passivity and equilibrium independent passivity (EIP) approaches, particularly for learning rules exhibiting imitation behavior, such as in replicator dynamics. To this end, we propose a method rooted in the concept of counterclockwise (CCW) dissipativity originally proposed to study stability and robustness of nonlinear feedback systems.

We proved that continuous memoryless payoff mechanisms are CCW if and only if they are potential games. Furthermore, we established that under (possibly dynamic) CCW payoff mechanisms, the population state converges to the Nash equilibria set for any learning rule within a convex cone spanned by imitation rules, continuous $\delta$-passive rules and any rule satisfying positive correlation. 

Hence, our work shows that CCW dissipativity offers a unified framework that encompasses both rules satisfying $\delta$-passivity and imitation-based rules. While our approach does not supersede existing methods in all cases, it significantly broadens the scope of cases that can be analyzed for Nash equilibrium learning. Our results also provide a theoretical foundation for future work seeking to design payoff mechanisms that ensure convergence to Nash equilibria, even when precise knowledge of the population's learning rule is lacking.

\bibliographystyle{elsarticle-num}        %
\bibliography{MartinsRefs,localref}

\end{document}